# Covalently Binding the Photosystem I to Carbon Nanotubes

S. Kaniber[1,3], L. Frolov[2], F.C. Simmel[3], A.W. Holleitner[1,3], C. Carmeli[2], and I. Carmeli[2]

*1 Walter Schottky Institut, Technical University Munich, Am Coulombwall 3, D-85748 Garching, Germany.*
*2 Department of Chemistry and Biochemistry, Tel-Aviv University, Tel-Aviv 69978, Israel.*
*3 Physik Department, Technical University Munich, James Franck Str., D-85748 Garching, Germany.*

**Abstract.** We present a chemical route to covalently couple the photosystem I (PS I) to carbon nanotubes (CNTs). Small linker molecules are used to connect the PS I to the CNTs. Hybrid systems, consisting of CNTs and the PS I, promise new photo-induced transport phenomena due to the outstanding electro-optical properties of the robust cyanobacteria membrane protein PS I.

## INTRODUCTION

Photosynthetic reaction centers are the photochemical active complexes in photosynthetic systems found in plants, algae and photosynthetic bacteria [1]. The photosynthetic reaction centers have evolved approximately 3.5 billion years ago, and they serve as the ultimate source of energy in the biosphere. The process involves an efficient conversion of solar energy to a stable chemical energy. Such is the reaction center photosystem I (PS I) which acts as a nanosize photodiode composed of a protein chlorophyll complex that utilizes light to generate a photopotential of 1.2 V with a quantum efficiency of 1 and an intrinsic energy conversion efficiency of 58% (47% of the total absorbed light) [1-5]. The PS I protein has a cylindrical shape with a diameter of about 15 nm and a height of 9 nm. It is intriguing to incorporate PS I into optoelectronic nanoscale circuits to exploit its outstanding optoelectronic properties [6].

Recently, we have demonstrated the possibility to covalently bind the PS I reaction center directly to gold surfaces [7] as well as indirectly via a small linker molecule to GaAs surfaces [8] and to carbon nanotubes [9]. To this end, amino acids in the extra membrane loops of the PS I facing the cytoplasmic side of the bacterial membrane (oxidizing side) were mutated to cysteines (Cys) [Fig. 1(a)]; enabling the formation of covalent bonds with a metal surface or a chemically functionalized GaAs surface. The Cys located at extra membranal loops of the protein do not have steric hindrance, when placed on a solid surface e.g. of a gold electrode or CNTs as shown here. The mutations D235C/Y634C were selected near the special chlorophyll pair P700 to allow close proximity between the reaction center and the CNTs [7]. Our self-assembly approach facilitates efficient electronic junctions and avoids disturbance in the function of the reaction center. The covalent attachment of the PS I through the Cys further ensures the structural stability of the self-assembled, oriented PS I. As demonstrated recently [7,8] a dry oriented monolayer of PS I assembled on gold electrodes and GaAs surfaces exhibits charge transfer between PS I and the solid state surface.

In Fig. 1(b) the molecular structure of the hybrid system is sketched. In [9] we reported on a four step chemical route utilizing sulfo-SMCC (sulfosuccinimidyl 4-[N-maleimidomethyl] cyclohex­ane-1-carboxylate) in order to covalently bind amine-reactivated CNTs to the PS I. As a result, the PS I is connected to the CNTs via the linker molecule $X_1$ [Fig. 1(c) and (b)]. Here, we report on utilizing sulfo-MBS (m-Maleimidobenzoyl-N-hydroxysulfosuccinimide ester), which results in a linker molecule $X_2$ [Fig. 1(d)] with an aromatic hydrocarbon instead of the cyclohexane in $X_1$.

To evaluate the degree of hybridization between the modified CNTs and PS I, a drop of the supernatant solution was placed onto a silicon surface, and incubated for two hours.

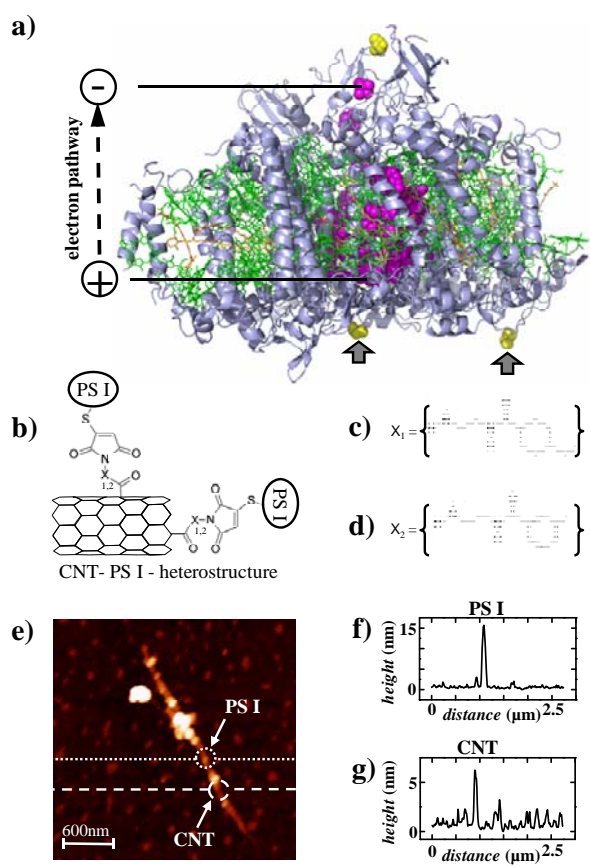

**FIGURE 1. (a)** Molecular structure image of the photosystem I (PS I) based upon crystallographic data. PS I is composed of polypeptide chains (gray) in which chlorophyll (green) and carotenoids (orange) are imbedded. The chromophores which mediate the electron transfer are represented by the space fill model (cyano). The PS I covalently binds to maleimide modified carbon nanotubes via Cys mutations along the polypeptide backbone [9]. In the present case, the Cys mutations of the mutant PS I were only located at the oxidizing (bottom) side of the PS I (see grey arrows). The dashed arrow schematically depicts the light induced charge separation across the PS I. **(b)** Sketch of the heterostructures made of a carbon nanotube and the PS I following the chemical scheme as in [9]. **(c)** and **(d)** Parts $X_1$ and $X_2$ of the linker molecules as in (b). **(e)** Atomic force micrograph of PS I bound to single-wall CNTs. The image shows a large number of PS I with a diameter of about 10-20 nm bound to the tips of CNTs and the side-walls (dotted circle). **(f)** Single trace along the dotted line in (e) showing the height of the PS I. **(g)** Single trace along the dashed line in (e) showing the height of the CNT bundle to be ~6 nm.

The samples were washed briefly with deionized water and dried under nitrogen. Figure 1(e) shows an atomic force micrograph (AFM) of the CNTs-PS I hybrid systems on a surface. The images exhibit a large number of spherical particles [dotted circle in Fig. 1(e)] attached to the surface of the CNTs. The height analysis of the AFM images [e.g. Fig. 1(f) and dotted line in Fig. 1(e)] indicates a height of about 9-20 nm for the spherical particles, in agreement with the actual diameter of the PS I, which suggests that the spherical particles are the PS I proteins. The diameter of the CNTs is in the range between 1 and 6 nm [Fig. 1(g) and dashed line in Fig. 1(e)]. In the particular case, a height of 6 nm suggests that a bundle of CNTs builds the back-bone of the hybrid system, consisting of CNTs and PS I.

## SUMMARY

In summary the present work demonstrates how to covalently couple carbon nanotubes (CNTs) to the photosynthetic reaction center I (PS I). In particular, we demonstrate that sulfo-MBS can be utilized to bind amine-activated CNTs to the PSI. Hybrid systems, consisting of CNTs and the PS I, promise new photo-induced transport phenomena, since the PS I is a robust cyanobacterial membrane protein with outstanding optical properties.

We gratefully acknowledge financial support by the DFG (Ho 3324/4), the Center for NanoScience (CeNS), the LMUexcellence program and the German excellence initiative via the "Nanosystems Initiative Munich (NIM)".